\documentclass[aps,prd,onecolumn,groupedaddress,showpacs,nofootinbib,amssymb
]{revtex4}
\usepackage[dvips]{graphicx}
\usepackage{amssymb}
\usepackage{amsmath}
\usepackage{graphicx}
\usepackage{amsfonts}
\usepackage{bm}
\usepackage{xcolor}
\begin{document}

\title{Finite-time Singularities in Swampland-related Dark Energy Models}
\author{S.D. Odintsov,$^{1,2,3}$\,\thanks{odintsov@ieec.uab.es}
V.K. Oikonomou,$^{4,5,6}$\,\thanks{v.k.oikonomou1979@gmail.com}}
\affiliation{$^{1)}$ ICREA, Passeig Luis Companys, 23, 08010 Barcelona, Spain\\
$^{2)}$ Institute of Space Sciences (ICE,CSIC) C. Can Magrans s/n,
08193 Barcelona, Spain\\
$^{3)}$ Institute of Space Sciences of Catalonia (IEEC),
Barcelona, Spain\\
$^{4)}$ Department of Physics, Aristotle University of Thessaloniki, Thessaloniki 54124, Greece\\
$^{5)}$ Laboratory for Theoretical Cosmology, Tomsk State University
of Control Systems
and Radioelectronics (TUSUR), 634050 Tomsk, Russia\\
$^{6)}$ Tomsk State Pedagogical University, 634061 Tomsk, Russia
}
\tolerance=5000

\begin{abstract}
In this work we shall investigate the singularity structure of the
phase space corresponding to an exponential quintessence dark
energy model recently related to swampland models. The dynamical
system corresponding to the cosmological system is an autonomous
polynomial dynamical system, and by using a mathematical theorem
we shall investigate whether finite-time singularities can occur
in the dynamical system variables. As we demonstrate, the
solutions of the dynamical system are non-singular for all cosmic
times and this result is general, meaning that the initial
conditions corresponding to the regular solutions, belong to a
general set of initial conditions and not to a limited set of
initial conditions. As we explain, a dynamical system singularity
is not directly related to a physical finite-time singularity.
Then, by assuming that the Hubble rate with functional form
$H(t)=f_1(t)+f_2(t)(t-t_s)^{\alpha}$, is a solution of the
dynamical system, we investigate the implications of the absence
of finite-time singularities in the dynamical system variables. As
we demonstrate, Big Rip and a Type IV singularities can always
occur if $\alpha<-1$ and $\alpha>2$ respectively. However, Type II
and Type III singularities cannot occur in the cosmological
system, if the Hubble rate we quoted is considered a solution of
the cosmological system.
\end{abstract}


\maketitle

\section{Introduction}

The consistent explanation of the late-time acceleration observed
in the late 90's \cite{Riess:1998cb}, dubbed dark energy, is a
theoretical challenge to modern cosmologists. Up to date, there
are various theoretical contexts that can explain this dark energy
era, for example modified gravity
\cite{reviews1,reviews2,reviews3,reviews4,reviews5,reviews6},
quintessence models
\cite{Carroll:1998zi,Zlatev:1998tr,Wang:1999fa,Barreiro:1999zs,Faraoni:2000wk,Gasperini:2001pc,Capozziello:2002rd,Sahni:2002kh,Capozziello:2003tk,Capozziello:2003gx,Caldwell:2005tm,Han:2018yrk}
and so on, and the challenge is to find a consistent with the
observational data description. Recently, the quintessence models
of dark energy were considerably constrained by the string theory
swampland criteria, firstly derived in Ref.
\cite{Vafa:2005ui,Ooguri:2006in}. Many works appeared after the
papers \cite{Vafa:2005ui,Ooguri:2006in}, addressing the swampland
criteria in the context of quintessence models of dark energy, see
for example
\cite{Obied:2018sgi,Agrawal:2018own,Heisenberg:2018yae,Akrami:2018ylq,Murayama:2018lie,Marsh:2018kub,Wang:2018kly}.
Basically, the swampland criteria constrain the field theories by
ruling out those which are incompatible with quantum gravity. The
implications of the swampland criteria on scalar models of dark
energy are also very serious, see for example
\cite{Heisenberg:2018yae} for a recent account on these works. Due
to the importance of the quintessence models for the dark energy
explanation, in this paper we shall study the singularity
structure of the phase space of exponential quintessence models.
The latter serve as a characteristic class of quintessence models,
and have been seriously constrained by the swampland criteria
\cite{Heisenberg:2018yae}. The singularity structure of the phase
space can provide insights for the occurrence of physical
singularities in the quintessence theory, as we shall demonstrate.
Indeed, a singularity in a phase space variable does not
necessarily imply that an actual physical singularity occurs, see
\cite{Odintsov:2018uaw} for an example of this sort in the context
of $f(R)$ gravity. In this work, we shall adopt the autonomous
dynamical system approach
\cite{Odintsov:2018uaw,Odintsov:2017tbc,Odintsov:2018awm,Kleidis:2018cdx},
and by using the dominant balances technique \cite{goriely}, we
shall reveal the finite-time singularity structure of the phase
space. For a recent review on dynamical systems applications in
cosmology see \cite{Bahamonde:2017ize}. Accordingly, we shall
investigate the implications of our findings on the singularity
structure of the physical theory. As we shall demonstrate based on
mathematical criteria, the phase space of the exponential
quintessence models is free of finite-time singularities, so in
view of this result, we question the possibility of having
physical finite-time singularities. By following the
classification of Ref. \cite{Nojiri:2005sx}, our results indicate
that Type II and Type III physical finite-time singularities
cannot occur in exponential quintessence models, however Big Rip
and Type IV singularities can occur. With our approach we provide
hints that Big Rip and Type IV are formally allowed to occur, for
a certain class of Hubble rate functional forms, however we
provide no actual proof that these types of singularities do
actually occur. This indicates that string cosmology may not be
free of future singularities like other quintessential dark
energy, if eventually a specific class of cosmological evolutions
is a solution to the cosmological equations. However we need to
note that in power-law potentials, geodesically complete future
singularities might actually occur
\cite{Lymperis:2017ulc}\footnote{In the end of the next section,
we shall discuss the issue of geodesics incompleteness and
spacetime singularities.}.  Let us note here that the relation
with the swampland criteria is quite strong since the exponential
potential corresponds to the most stable attractor solution of the
swampland general phase space \cite{Agrawal:2018own}. In the
terminology and notation of Ref. \cite{Agrawal:2018own} this
stable attractor corresponds to the constant $\lambda$ case, so
our study essentially focuses on properties of this specific
attractor solution.

This paper is organized as follows: In section II we briefly
review the essential features of quintessence models and we
present the dynamical system corresponding to exponential
quintessence models. In section III we investigate the singularity
structure of the phase space of exponential quintessence models
and we discuss the implications of our results on the physical
singularity structure of the quintessence models under study.
Finally, the conclusions follow in the end of the paper.

Before we start, we discuss in brief the geometric conventions
which we shall use in this work. Particularly, we shall assume
that the metric is a flat Friedmann-Robertson-Walker (FRW) of the
form,
\begin{equation}\label{frw}
ds^2 = - dt^2 + a(t)^2 \sum_{i=1,2,3} \left(dx^i\right)^2\, ,
\end{equation}
where $a(t)$ is the scale factor of the Universe. In addition, we
shall use a physical units system in which $\hbar=c=1$.

\section{Essential Features of the Exponential Quintessence Dark Energy Models and their Dynamical System}

The action of general scalar field quintessence models in vacuum
for a general metric $g_{\mu \nu}$ has the following form,
\begin{equation}\label{generalactionforquintessence}
\mathcal{S}=\int
d^{4}x\sqrt{-g}\left(\frac{1}{\kappa^2}R-\frac{1}{2}\partial_{\mu}\phi\partial^{\mu}\phi-V(\phi)
\right)\, ,
\end{equation}
where $\kappa^2=\frac{1}{M_p}$ and $M_p$ is the Planck mass scale.
The equations of motion for this scalar theory in the FRW
background of Eq. (\ref{frw}) read,
\begin{align}\label{scalarequationsofmotion}
& \ddot{\phi}+3H\dot{\phi}+V'(\phi)=0\, ,\\ \notag &
\frac{3H^2}{\kappa^2}=\frac{\dot{\phi}^2}{2}+V(\phi)\, ,
\end{align}
where the prime denotes differentiation with respect to the scalar
field $\phi$. We shall focus on exponential quintessence models,
so the scalar potential $V(\phi)$ has the form
$V(\phi)=e^{-\frac{\lambda \phi}{M_p}}$, in which case the
equations of motion (\ref{scalarequationsofmotion}) can be cast in
the form of an autonomous dynamical system. This can be achieved
by using the following variables,
\begin{equation}\label{variablesdynamicalsystem}
x=\frac{\dot{\phi}}{\sqrt{6}H\,M_p},\,\,\,y=\frac{\sqrt{V}}{\sqrt{3}H\,M_p}\,
,
\end{equation}
where the ``dot'' indicates differentiation with respect to the
cosmic-time $t$. By using the $e$-foldings number $N$ as a
dynamical variable, we can construct an autonomous dynamical
system by combining Eqs. (\ref{scalarequationsofmotion}) and
(\ref{variablesdynamicalsystem}) which has the following form,
\begin{align}\label{dynamicalsystem}
& \frac{\mathrm{d}x_1}{\mathrm{d}N}=-3x+\frac{\sqrt{6}}{2}\lambda
y^2+\frac{3}{2}x\left(x^2-y^2\right) \, ,
\\ \notag &
\frac{\mathrm{d}x_2}{\mathrm{d}N}=-\frac{\sqrt{6}}{2}\lambda x
y+\frac{3}{2}y\left(x^2-y^2\right)\, .
\end{align}
Also due to the second equation in Eq.
(\ref{scalarequationsofmotion}), the variables $x$ and $y$ satisfy
the Friedmann constraint,
\begin{equation}\label{friedmanconstraint}
x^2+y^2=1\, ,
\end{equation}
which must be respected even at a finite-time singularity. This is
a well-known feature in dynamical system analysis, se for example
Ref. \cite{Odintsov:2018uaw}, where the singular solutions satisfy
the Friedmann constraint due to the fact that the singularities
cancel. The above dynamical system is an autonomous dynamical
system of polynomial form. We shall not focus on the phase space
structure of this dynamical system, since the fixed points and
their stability were studied in Ref. \cite{Agrawal:2018own}, but
we mainly focus on the singularity structure of this dynamical
system. This is the subject of the next section.

\section{Singularity Occurrence in Swampland Dark Energy Models: Dominant Balance Analysis and Physical Interpretation}

The occurrence or not of finite-time singularities in polynomial
autonomous dynamical systems was studied some time ago in Ref.
\cite{goriely}. The authors of Ref. \cite{goriely} developed an
asymptotic analysis technique which we shall call dominant balance
analysis hereafter, which can determine whether general singular
or non-singular solutions exist in the dynamical system. This
analysis was applied in cosmological systems in Ref.
\cite{barrowcotsakis}, see also Refs.
\cite{Odintsov:2018uaw,Odintsov:2018awm} for some recent
applications. In order to maintain the article self-contained and
coherent we shall briefly review this dominant balance technique,
and we directly apply this for the autonomous dynamical system of
Eq. (\ref{dynamicalsystem}).

Consider the following dynamical system,

\begin{equation}\label{dynamicalsystemdombalanceintro}
\dot{x}=f(x)\, ,
\end{equation}
where the $n$-dimensional vector $x$ is $(x_1,x_2,...,x_n)$, and
$f(x)$ is a $n$-dimensional vector of the form
$f(x)=\left(f_1(x),f_2(x),...,f_n(x)\right )$. We assume that the
functions $f_i(x)$ are polynomials of the dynamical system
variables $(x_1,x_2,...,x_n)$. At a finite-time singularity
$t=t_c$, some variables of the dynamical system will have the form
$(t-t_c)^{-p}$ with $p$ being some positive number (or
equivalently in terms of $N$, $(N-N_c)^{-q}$, with $q\neq p$ in
general). Having the above assumptions in mind, the method of the
dominant balance analysis has the following steps:

\begin{itemize}

\item Determine all the possible truncations of the vector
function $f(x)$ appearing in Eq.
(\ref{dynamicalsystemdombalanceintro}). One of these truncations
will dominate the evolution near a finite-time singularity, so we
denote this truncation as $\hat{f}(x)$, and hence the dynamical
system near the singularity takes the form,
\begin{equation}\label{dominantdynamicalsystem}
\dot{x}=\hat{f}(x)\, .
\end{equation}
After that, we write
\begin{equation}\label{decompositionofxi}
x_1(\tau)=a_1\tau^{p_1},\,\,\,x_2(\tau)=a_2\tau^{p_2},\,\,\,....,x_n(\tau)=a_n\tau^{p_n}\,
,
\end{equation}
where $\tau=t-t_c$, hence we made the assumption that the vector
$x$ is written in $\psi$-series near the singularity. By
substituting Eq. (\ref{decompositionofxi}) in Eq.
(\ref{dominantdynamicalsystem}), and equating the exponents we get
the numbers $p_i$, $i=1,2,...,n$ which must be real fractional or
integer numbers. With the $p_i$'s at hand, we form the vector
$\vec{p}=(p_1,p_2,...,p_n)$, and by substituting these in
(\ref{dominantdynamicalsystem}), (\ref{decompositionofxi}), by
equating the polynomials, we finally determine the numbers $a_i$
appearing in Eq. (\ref{decompositionofxi}), and we form the vector
$\vec{a}=(a_1,a_2,a_3,....,a_n)$. The vectors
$(\vec{a},\vec{p})\neq 0$ constitute a dominant balance in the
terminology of Ref. \cite{goriely}.

\item The next steps of the theorem-method of dominant balances is
easy to comprehend. If the vector $\vec{a}=(a_1,a_2,a_3,....,a_n)$
has complex entries, then the theorem of Ref. \cite{goriely}
clearly states that no finite-time singularities occur in the
dynamical system (\ref{dynamicalsystemdombalanceintro}), but if it
has real entries, then the dynamical system has finite-time
singularities, and some of it's variables $x_i$, blow up.

\item At the next step, one must ensure that the solutions found
in the previous step are general or non-general. Let us clarify
that a general solution corresponds to a general set of initial
conditions, while a non-general solution corresponds to a limited
set of initial conditions. To this end, we construct the
Kovalevskaya matrix $R$, which is,
\begin{equation}\label{kovaleskaya}
R=\left(%
\begin{array}{ccccc}
  \frac{\partial \hat{f}_1}{\partial x_1} & \frac{\partial \hat{f}_1}{\partial x_2} & \frac{\partial \hat{f}_1}{\partial x_3} & ... & \frac{\partial \hat{f}_1}{\partial x_n} \\
  \frac{\partial \hat{f}_2}{\partial x_1} & \frac{\partial \hat{f}_2}{\partial x_2} & \frac{\partial \hat{f}_2}{\partial x_3} & ... & \frac{\partial \hat{f}_2}{\partial x_n} \\
  \frac{\partial \hat{f}_3}{\partial x_1} & \frac{\partial \hat{f}_3}{\partial x_2} & \frac{\partial \hat{f}_3}{\partial x_3} & ... & \frac{\partial \hat{f}_3}{\partial x_n} \\
  \vdots & \vdots & \vdots & \ddots & \vdots \\
  \frac{\partial \hat{f}_n}{\partial x_1} & \frac{\partial \hat{f}_n}{\partial x_2} & \frac{\partial \hat{f}_n}{\partial x_3} & ... & \frac{\partial \hat{f}_n}{\partial x_n} \\
\end{array}%
\right)-\left(%
\begin{array}{ccccc}
  p_1 & 0 & 0 & \cdots & 0 \\
  0 & p_2 & 0 & \cdots & 0 \\
  0 & 0 & p_3 & \cdots & 0 \\
  \vdots & \vdots & \vdots & \ddots & 0 \\
  0 & 0 & 0 & \cdots & p_n \\
\end{array}%
\right)\, .
\end{equation}
Next we calculate the Kovalevskaya matrix $R$ at a non-zero
balance found at a previous step, and we calculate the
eigenvalues. If the method is applied correctly, a consistent
resulting Kovalevskaya matrix $R$ will have eigenvalues of the
form $(-1,r_2,r_3,...,r_{n})$.

\item If in the previous step we find $r_i>0$, $i=2,3,...,n$, then
the solutions we found are general, but if one of the above
eigenvalues is negative, then the solutions we found at a previous
step correspond to a limited set of initial conditions and are
thus non-general.

\end{itemize}

Let us apply the method we described above, for the dynamical
system (\ref{dynamicalsystem}), subject to the Friedmann
constraint (\ref{friedmanconstraint}). To this end, we rewrite the
dynamical system (\ref{dynamicalsystem}), in the form,
\begin{equation}\label{dynsysnewform}
\frac{\mathrm{d}\vec{x}}{\mathrm{d}N}=f(\vec{x})\, ,
\end{equation}
where $\vec{x}=(x,y)$, and also the vector-valued function
$f(\vec{x})$ is equal to,
\begin{equation}\label{functionfmultifluidclassicalcase}
f(x,y)=\left(%
\begin{array}{c}
 f_1(x,y) \\
  f_2(x,y) \\
\end{array}%
\right)\, ,
\end{equation}
where the $f_i$'s in Eq. (\ref{functionfmultifluidclassicalcase})
are equal to,
\begin{align}\label{functionsficlassicalcase}
& f_1(x,y)=-3x+\frac{\sqrt{6}}{2}\lambda
y^2+\frac{3}{2}x\left(x^2-y^2\right)\, ,
\\
\notag & f_2(x,y)=-\frac{\sqrt{6}}{2}\lambda x
y+\frac{3}{2}y\left(x^2-y^2\right)
 \, .
\end{align}
There exist several truncations of the vector-valued function
$f(\vec{x})$, but a consistent truncation is the following,
\begin{equation}\label{truncation1classicalcase}
\hat{f}(x,y)=\left(
\begin{array}{c}
 -\frac{3 x y^2}{2} \\
\frac{3 x^2 y}{2} \\
\end{array}
\right)\, .
\end{equation}
By using the method we presented previously, we easily find that
the corresponding vector $\vec{p}$ has the form,
\begin{equation}\label{vecp1classicalcase}
\vec{p}=( -\frac{1}{2},-\frac{1}{2} )\, ,
\end{equation}
and in the same way, the following non-zero vector solutions
$\vec{a}$ are obtained,
\begin{align}\label{balancesdetails1classicalcase}
& \vec{a}_1=\Big{(}-\frac{i}{\sqrt{3}}, -\frac{1}{\sqrt{3}}\Big{)}
\\ \notag & \vec{a}_2=\Big{(}-\frac{i}{\sqrt{3}}, \frac{1}{\sqrt{3}}\Big{)}\, ,
\\ \notag & \vec{a}_3=\Big{(}\frac{i}{\sqrt{3}}, -\frac{1}{\sqrt{3}}\Big{)}\, ,
\\ \notag &
\vec{a}_4=\Big{(}\frac{i}{\sqrt{3}}, \frac{1}{\sqrt{3}}\Big{)}\, .
\end{align}
As it is obvious, all the solutions $\vec{a}_i$, $i=1,..4$ are
complex, and also it can be easily checked that Friedman
constraint (\ref{friedmanconstraint}) can be satisfied for all the
solutions $\vec{a}_i$ and for the vector $\vec{p}$ being chosen as
in Eq. (\ref{vecp1classicalcase}). This can easily be seen, since
the expression $x^2+y^2$ appearing in the Friedmann constraint,
for all the $\vec{a}_i$ reads at leading order (only the leading
order terms are shown),
\begin{equation}\label{friedmannconstraintatasingularity}
\left(\pm \frac{i}{\sqrt{3}}\right)^2\tau^{-\frac{1}{2}}+\left(\pm
\frac{1}{\sqrt{3}}\right)^2\tau^{-\frac{1}{2}}\, ,
\end{equation}
which is equal to zero, where $\tau=N-N_c$. Having verified that
the Friedmann constraint can never become singular, let us proceed
to the problem at hand, and with regard to the question whether
singular solutions exist, the answer is no, since the solutions
$\vec{a}_i$ are complex. Now what remains is to check whether
these solutions are general or not. To this end, we shall
calculate the corresponding Kovalevskaya matrix $R$, which in our
case  is,
\begin{equation}\label{kovalev1classicalcase}
R=\left(
\begin{array}{cc}
 \frac{1}{2}-\frac{3 y^2}{2} & -3 x y \\
 3 x y & \frac{3 x^2}{2}+\frac{1}{2} \\
\end{array}
\right)\, .
\end{equation}
By evaluating the Kovalevskaya matrix $R$ on the solution
$(x,y)=\vec{a}_1$, we obtain the matrix,
\begin{equation}\label{ra}
R(\vec{a})=\left(
\begin{array}{cc}
 0 & -i \\
 i & 0 \\
\end{array}
\right)\, ,
\end{equation}
and by calculating the corresponding eigenvalues we get,
\begin{equation}\label{eigenvalues1classicalcase}
(r_1,r_2)=(-1,1)\, .
\end{equation}
The same eigenvalues as above can be found for all the solutions
$\vec{a}_i$, $i=2,3,4$. Hence, in view of the theorem we presented
above, the dynamical system (\ref{dynamicalsystem}) never develops
finite-time singularities, and these non-singular solutions are
general, which means that these correspond to a general set of
initial conditions, and not to a limited set of initial
conditions. We must note that there is another non-trivial
consistent truncation, however the form of the resulting $\vec{p}$
makes the Friedmann constraint to leading order peculiarities, so
this case is excluded from our analysis.

Let us now investigate what the results we found indicate for the
physical finite-time singularity structure of the cosmological
system at hand. Basically, the variables $x$ and $y$ can never
become singular, so this constraint restricts the allowed types of
singularities that can occur. Let us recall in brief the
classification of finite-time singularities according to Ref.
\cite{Nojiri:2005sx}, so there are four types of singularities at
$t=t_s$:
\begin{itemize}
\item Type I (``Big Rip''): This is the most severe type of
singularity, from a phenomenological point of view, and in this
case, as $t \to t_s$, $a \to \infty$, $\rho_\mathrm{eff} \to
\infty$ and $\left|p_\mathrm{eff}\right| \to \infty$.

 \item Type II
(``sudden''): This is known as pressure singularity, and in this
case, as $t \to t_s$, $a \to a_s$, $\rho_\mathrm{eff} \to \rho_s$
and $\left|p_\mathrm{eff}\right| \to \infty$.

\item Type III: In this case as $t \to t_s$, $a \to a_s$,
$\rho_\mathrm{eff} \to \infty$ and $\left|p_\mathrm{eff}\right|
\to \infty$.

 \item Type IV : This is a soft type of singularity,
and in this case, as $t \to t_s$, $a \to a_s$, $\rho_\mathrm{eff}
\to \rho_s$, $\left|p_\mathrm{eff}\right| \to p_s$ and the higher
derivatives ($n>2$) of $H$ diverge.
\end{itemize}
In the above classification, $\rho_\mathrm{eff}$ and
$p_\mathrm{eff}$ stand for the total energy density and the total
pressure of the cosmological system. The most general form of a
Hubble rate developing some of the above listed singularities is
the following,
\begin{equation}\label{hubblerategeneral}
H(t)=f_1(t)+f_2(t)(t-t_s)^{\alpha}\, ,
\end{equation}
where for consistency we require the parameter $\alpha$ to be
$\alpha=\frac{2n}{2m+1}$, where $m,n$ positive integers. The
functions $f_1(t)$ and $f_2(t)$ are required to be regular at the
singularity time instance $t=t_s$ and also $f_1(t_s)\neq 0$,
$f_2(t_s)\neq 0$. Also their derivatives up to second order are
assumed to satisfy the same constraints. The Hubble rate
(\ref{hubblerategeneral}) is not necessarily a solution to the
field equations, but we will discuss the case that indeed it is a
solution, and we will investigate the implications of such a
solution on the singularity structure, in view of the results we
found for the dynamical system. Particularly, the variables $x\sim
\frac{\dot{\phi}}{H}$ and $y\sim \frac{\sqrt{V(\phi)}}{H}$ must be
non singular for all the values of the cosmic time. The various
singularities that the Hubble rate (\ref{hubblerategeneral}) can
generate, depending on the values of the parameter $\alpha$ are
listed below,
\begin{itemize}
    \item If $\alpha<-1$ a Big Rip singularity occurs.
    \item If $-1<\alpha < 0$ a Type III singularity occurs.
    \item If $0<\alpha < 1$ a Type II singularity occurs.
    \item If $\alpha > 1$ a Type IV singularity occurs.
\end{itemize}
So let us express the terms $\dot{\phi}$ and $V(\phi)$ in terms of
the Hubble rate, so by combining Eqs.
(\ref{scalarequationsofmotion}), we get,
\begin{equation}\label{neweqnmotion}
\frac{2}{\kappa^2}\dot{H}=-\dot{\phi}^2\, ,
\end{equation}
so the above can be solved in terms of $\dot{\phi}$ and the
variable $x\sim \frac{\dot{\phi}}{H}$ can be expressed in terms of
the Hubble rate in closed form. Also, by using the fact that
$V'(\phi)=-\lambda V(\phi)$ for the exponential potential under
study, we obtain the potential as a function of the Hubble rate,
by combining Eqs. (\ref{scalarequationsofmotion}) and
(\ref{neweqnmotion}), so we finally have,
\begin{equation}\label{potentialasfunctionof}
V(\phi)=-\frac{\ddot{H}+6\dot{H}}{\kappa^2\lambda \dot{\phi}}\, ,
\end{equation}
and in effect, the potential $V(\phi)$ is also expressed in closed
form as a function of the Hubble rate, in view also of Eq.
(\ref{neweqnmotion}). Without presenting the resulting
expressions, which are quite lengthy, let us investigate which
singularities can occur, in view of the fact that the variables
$x$ and $y$ are always regular. By studying the resulting
expressions, for the Type IV singularity case, the only case that
this singularity cannot occur is when $1<\alpha <2$, due to the
presence of a term $\sim (t-t_s)^{\alpha-2}$ in the final
expression of the variable $y$. However, when $\alpha>2$, the
variables $x$, $y$ are always regular, so a Type IV singularity
can always occur for $\alpha>2$. For the Type II case, which
occurs for $0<\alpha<1$, the variable $x$ is always singular due
to the presence of a term $\sim (t-t_s)^{\alpha-1}$, so it cannot
occur at all. With regard to the Type III case, it cannot occur
due to the presence of a term $\sim
(t-t_s)^{-\frac{3\alpha}{2}-\frac{3}{2}}$, which is always
singular for $0>\alpha>-1$. Finally, the Big Rip singularity can
always occur, since for $\alpha<-1$ the variables $x$ and $y$ are
always regular. We gathered the results of our investigation in
Table \ref{table1}.
\begin{table*}[h]
\small \caption{\label{table1}Allowed Singularity Types for the
Hubble rate $H(t)=f_1(t)+f_2(t)(t-t_s)^{\alpha}$}
\begin{tabular}{@{}crrrrrrrrrrr@{}}
\tableline \tableline \tableline
 Type I (Big Rip) &$\,\,\,\,$ Always occurs when $\alpha<-1$.
\\\tableline
Type II: & Cannot occur \\
\\\tableline
Type III: & Cannot occur. \\
Type IV: & Cannot occur for $1<\alpha<2$, but can occur for $\alpha>2$.\\
\\\tableline
 \tableline
\end{tabular}
\end{table*}
In conclusion, if the Hubble rate (\ref{hubblerategeneral}) is a
solution of the field equations, the only Types of singularities
that can never occur are the Type III, and the Type II. The other
two types of singularities, namely the Type IV and Type I can
always occur when $\alpha>2$ for the Type IV, and for any
$\alpha<-1$ for the Big Rip case.

 Before closing let us briefly clarify the meaning of
a spacetime singularity. In the end of the introduction we
mentioned that for certain future finite-time singularities, like
for example the sudden singularities, no geodesics incompleteness
occurs. Actually, for a strong singularity, geodesics
incompleteness always occurs, so for soft singularities, the
terminology singularity is used just to indicate the divergence of
certain physical quantities, or of certain higher curvature
invariants. According to the classification of Ref.
\cite{Nojiri:2005sx} which we presented in the list above Eq.
(\ref{neweqnmotion}), finite-time singularities occur when the
physical quantities pressure and energy density, or the scale
factor diverge. However let us here discuss in some detail the
issue of spacetime singularity. In our opinion, Penrose's
definition of a singularity is the most accurate, according to if
appropriate energy conditions are satisfied, a singular spacetime
point is equivalent with geodesics incompleteness. This is the
stronger definition and most accurate definition of the occurrence
of a singularity. This also covers the case of a flat spacetime
with a point removed, which is not geodesically complete and it is
singular, in spite of having null curvature in all its points.
This brings us to the second trend in the literature, on how
singularities are defined. Specifically, the occurrence of
spacetime singularities in smooth manifolds is accompanied with
curvature divergences, and specifically integrals of the higher
curvature invariants along a geodesic which meets a strong
singularity at some finite-time $t_s$, like the following which
uses the Riemann tensor \cite{FernandezJambrina:2004yy},
\begin{equation}\label{geodesicincompleteness}
\int_{0}^{t_s}d \tau'\int_0^{\tau'}d\tau''R^{i}_{0j0}(\tau'')\, ,
\end{equation}
strongly diverge at the time instance. So in our opinion, the best
definition of a singularity means geodesic incompleteness. This
geodesic incompleteness may effectively bring along higher
curvature divergences, however the definition of curvature in an
open set around the spacetime singularity may be considered ill
defined. In fact, a speculation of what a strong finite-time
singularity might mean for the spacetime, was given in Ref.
\cite{Oikonomou:2018qsc}, where it was claimed that the topology
of the spacetime actually changes when the singularity occurs. In
such contexts, a curvature invariant is ill defined, since the
differentiability of the spacetime is lost, it is not smooth
anymore. Let us note that the issue of defining a singular
spacetime point is one of the most difficult issues in general
relativity. Actually, Penrose himself could not accept the Big
Bang singularity itself, quoting that it is not a spacetime point
but an initial singular hypersurface \cite{penrosenontrivial}.
This is due to the fact that a spacetime point singularity would
mean an infinity of overlapping particle horizons, which
eventually would lead to an infinite number of causally
disconnected regions in an expanding Universe. We will thus not
further discuss this highly non-trivial issue, which extends by
far the purposes of this article.

\section{Conclusions}

In this paper we studied the singularity structure of the phase
space corresponding to an exponential quintessential model of dark
energy, which is constrained by the swampland criteria.
 The exponential model corresponds to the most stable
attractor solution of the swampland scalar field phase space
\cite{Agrawal:2018own}, so our study mainly focuses on the
properties of this attractor solution. We focused on the question
whether the variables of the dynamical system can become singular
at some finite-time instance, and due to the fact that the
dynamical system is polynomial, we applied a mathematical theorem
in order to see whether finite-time singularities can occur in the
dynamical system variables. As we demonstrated, the variables of
the dynamical system can never become singular, so this imposes
restrictions on the physical finite-time singularities that the
cosmological system can develop. The result indicates actually
that the solutions of the dynamical system are always regular and
these correspond to a general set of initial conditions and not to
a limited set of initial conditions. We also clarified the
difference between a finite-time singularity in a dynamical system
variable and a physical finite-time singularity. As we
demonstrated, if the general Hubble rate
$H(t)=f_1(t)+f_2(t)(t-t_s)^{\alpha}$ is a solution of the
cosmological equations, then only Type IV and Big Rip
singularities can always occur for $\alpha>2$ and $\alpha<-1$
respectively. The Type II and Type III singularities cannot occur
for the cosmological system, if the Hubble rate we quoted above is
considered a solution of the system. What we did not study is the
effect of matter and radiation perfect fluids in the dynamical
system. In principle the singularity structure might change,
possibly the generality of solutions, however one should check
this in detail, and we hope to address this issue in a future
work.

\section*{Acknowledgments}

This work is supported by MINECO (Spain), FIS2016-76363-P, by
project 2017 SGR247 (AGAUR, Catalonia) (S.D. Odintsov) and by
Russian Ministry of Science and High Education, project No.
3.1386.2017.

\end{document}